\documentclass[aps,prl,twocolumn,superscriptaddress,showpacs]{revtex4}
\usepackage{graphicx}
\usepackage[latin1]{inputenc}
\usepackage{textcomp}
\usepackage{mathptmx}
\newcommand{\rb}{$^{87}$Rb}
\newcommand{\kt}{$^{39}$K}

\newcommand{\kq}{$^{41}$K}
\newcommand{\alphak}{\alpha_{\rm KKRb}}
\newcommand{\nk}{n_{\rm K}}
\newcommand{\Nk}{N_{\rm K}}
\newcommand{\mk}{m_{\rm K}}
\newcommand{\omegak}{\omega_{\rm K}}

\newcommand{\alpharb}{\alpha_{\rm KRbRb}}

\newcommand{\nrb}{n_{\rm Rb}}
\newcommand{\Nrb}{N_{\rm Rb}}
\newcommand{\mrb}{m_{\rm Rb}}
\newcommand{\omegarb}{\omega_{\rm Rb}}

\newcommand{\kB}{k_{\rm B}}
\begin{document}

\title{Observation of heteronuclear atomic Efimov resonances}

\author{G.~Barontini}
\affiliation{LENS, European Laboratory for Non-Linear Spectroscopy and
  Dipartimento di Fisica, Università di Firenze, via N. Carrara 1,
  I-50019 Sesto Fiorentino, Firenze, Italy}

\author{C.~Weber}%
\affiliation{Institut für Angewandte Physik, Universität Bonn,
  Wegelerstraße 8, D-53115 Bonn, Germany}

\author{F.~Rabatti}%
\affiliation{LENS, European Laboratory for Non-Linear Spectroscopy and
  Dipartimento di Fisica, Università di Firenze, via N. Carrara 1,
  I-50019 Sesto Fiorentino, Firenze, Italy}

\author{J.~Catani}%
\affiliation{LENS, European Laboratory for Non-Linear Spectroscopy and
  Dipartimento di Fisica, Università di Firenze, via N. Carrara 1,
  I-50019 Sesto Fiorentino, Firenze, Italy}%
\affiliation{CNR-INFM, via G. Sansone 1, I-50019 Sesto Fiorentino,
  Firenze, Italy}

\author{G.~Thalhammer}%
\affiliation{LENS, European Laboratory for Non-Linear Spectroscopy and
  Dipartimento di Fisica, Università di Firenze, via N. Carrara 1,
  I-50019 Sesto Fiorentino, Firenze, Italy}%

\author{M.~Inguscio}
\affiliation{LENS, European Laboratory for Non-Linear Spectroscopy and
  Dipartimento di Fisica, Università di Firenze, via N. Carrara 1,
  I-50019 Sesto Fiorentino, Firenze, Italy}%

\author{F.~Minardi}%
\affiliation{LENS, European Laboratory for Non-Linear Spectroscopy and
  Dipartimento di Fisica, Università di Firenze, via N. Carrara 1,
  I-50019 Sesto Fiorentino, Firenze, Italy}%
\affiliation{CNR-INFM, via G. Sansone 1, I-50019 Sesto Fiorentino,
  Firenze, Italy}

\date{\today}

\begin{abstract}
  The Efimov effect represents a cornerstone in few-body
  physics. Building on the recent experimental observation with
  ultracold atoms, we report the first experimental signature of
  Efimov physics in a heteronuclear system. A mixture of
  \kq\ and \rb\ atoms was cooled to
  few hundred nanoKelvins and stored in an optical dipole
  trap. Exploiting a broad interspecies Feshbach resonance, the losses
  due to three-body collisions were studied as a function of the
  interspecies scattering length. We observe an enhancement of the
  three-body collisions for three distinct values of the interspecies
  scattering lengths, both positive and negative. We attribute the two
  features at negative scattering length to the existence of
  two kind of Efimov trimers, namely KKRb and KRbRb.
\end{abstract}

\pacs{34.50.-s, 36.40.-c, 21.45.-v, 67.85.-d}





\maketitle

Sun-Earth-Moon, the Helium atom, the proton: at all length scales,
three-body systems are ubiquitous in physics, yet they challenge our
understanding in many ways. Their complexity conspicuously exceeds the
two-body counterparts. A peculiar class of three-body systems defying
our intuition arises when the constituents feature resonant pair-wise
interactions, such that the scattering length is much larger than the
effective range of the pair potential. In a few seminal papers,
V. Efimov advanced our understanding of such three-body systems and
demonstrated the existence of a large number of weakly bound
three-body states, thereafter known as the \emph{Efimov
  effect}\cite{Efimov1970, Efimov1971}. What makes Efimov states truly
remarkable is their \emph{universality}, i.e., the fact that their
main properties are independent from the details of the pair
potential, be it the strong interaction between two nucleons or the
van der Waals force between two neutral atoms.

For over 35 years, the Efimov effect sparked an intense theoretical
research \cite{Braaten2006}, while eluding experimental
observation. The first experimental evidence of Efimov states was only
recently reached with ultracold $^{133}$Cs \cite{Kraemer2006} and \kt\
\cite{Zaccanti2009} atoms, thanks to the possibility to adjust at will
the scattering length by means of Feshbach resonances. In nuclear
physics, the original context studied by V.~Efimov, the Efimov effect
is hampered by the strong long-range Coulomb interactions and
therefore confined to triads where at least two constituents are
neutral. Among these, halo nuclei, i.e., nuclei like $^6$He,
$^{11}$Li, $^{14}$Be, $^{20}$C composed of a smaller core nucleus plus
two loosely bound neutrons, have been identified as possible examples
of Efimov physics \cite{Jensen2004} and there is an ongoing debate
about the prospects of observing nuclear Efimov states
\cite{Mazumdar2006}. To this goal, it is crucial to study Efimov
physics in systems composed of distinguishable particles with different
masses.

In this work, we report the first experimental evidence of Efimov
physics with particles of different masses, i.e., Efimov resonances in
the three-body collisions of a mixture of ultracold \kq\ and \rb\
atoms. Our experiment demonstrates that two resonantly interacting
pairs are sufficient to grant the existence of Efimov
states\cite{Braaten2006} and, thanks to universality, suggests that
they could be observed also in other asymmetric triads, like the halo
nuclei.

In ultracold atomic gases, Efimov states fragile and unstable to decay
toward deeply bound two-body molecular levels. While direct
observation has never been achieved, the presence of Efimov states is
revealed by measuring the atomic losses, since it bears a dramatic
impact in the three-body recombination (3BR) collisions
\cite{Braaten2006, Kraemer2006}.

\begin{figure}
  \includegraphics[width=\columnwidth]{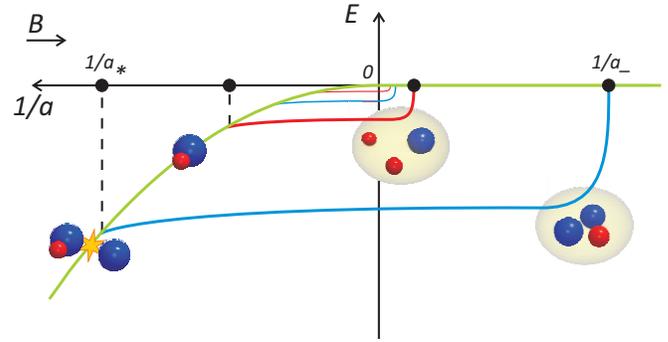}
  \caption{(Color online) Schematic energy diagram of Efimov states for our
    double-species mixture of K and Rb, around an interspecies
    Feshbach resonance where the K-Rb scattering length diverges
    ($1/a=0$). The Efimov states appear: (i) at the atom-dimer
    threshold for positive scattering lengths $a_\ast$; (ii) at the
    three-atoms threshold for negative scattering lengths $a_-$. Two
    distinct kind of Efimov trimer are possible, KKRb and KRbRb, shown
    respectively by red and blue lines. The green line shows the
    dissociation threshold of the Efimov states.}
  \label{fig:fig1}
\end{figure}
The energy of Efimov states depends on the resonant scattering length,
as depicted in Fig.~\ref{fig:fig1}. For a certain negative value of
the two body scattering length $a = a_-$, the binding energy of an
Efimov state vanishes, i.e., the energy of the trimer coincides with
that of three free atoms. At this scattering length $a_-$, the 3BR
collisions are resonantly enhanced. For positive values of the
scattering length, the Efimov scenario is richer. The 3BR rate
displays an oscillatory behavior with broad maxima and minima. In
addition, at a scattering length value $a = a_{\ast}$, the Efimov
trimers energy hits the atom-dimer threshold: here, a resonant
enhancement occurs for the atom-dimer collisions, both elastic and
inelastic. In the limit of infinite $|a|$, the above features repeat
for each state of the Efimov spectrum as the scattering length is multiplied
by integer powers of a scaling factor, usually denoted with
$e^{\pi/s_o}$. While the values of the resonant scattering lengths
$a_\ast$ and $a_-$ depend on the details of the atomic potential and
are so far unpredictable, theoretical predictions are available for
their ratios $a_\ast/a_-$, at least in systems of identical particles
\cite{Braaten2006, Helfrich2009}.

With the mixture of \kq\ and \rb, we have two distinct three-body loss
channels enhanced by the proximity of the interspecies Feshbach
resonance, namely KKRb and KRbRb collisions, as shown in
Fig.~\ref{fig:fig1}.  Correspondingly, there exist two Efimov series,
whose relative location is so far unpredictable, with different
scaling factors $e^{\pi/s_0}=3.51\times 10^5$ for KKRb and $131$ for
KRbRb \cite{Jensen2003}.

We now briefly describe our experimental procedure. We prepare the
ultracold atomic mixture by sympathetic cooling \cite{Modugno2001},
first in a magnetic trap and then in a crossed dipole trap, far
off-resonant with respect to both K and Rb atomic transitions
($\lambda=1064$\,nm). We cool the mixture to temperatures as low as
300\,nK, while trapping it in a harmonic potential ($\omegarb\simeq
2\pi\times 70$\,Hz, $\omegak=\omegarb\sqrt{\mrb/\mk}$). Both species
are prepared in the $|F = 1, m_F=1\rangle$ states, featuring a broad
interspecies Feshbach resonance at 38.4\,G \cite{Thalhammer2008},
which allows to adjust the interspecies scattering length $a$. Since
the $|1,1\rangle$ state is the absolute ground state, inelastic two
body collisions are suppressed.

\begin{figure}
  \includegraphics[width=.9\columnwidth]{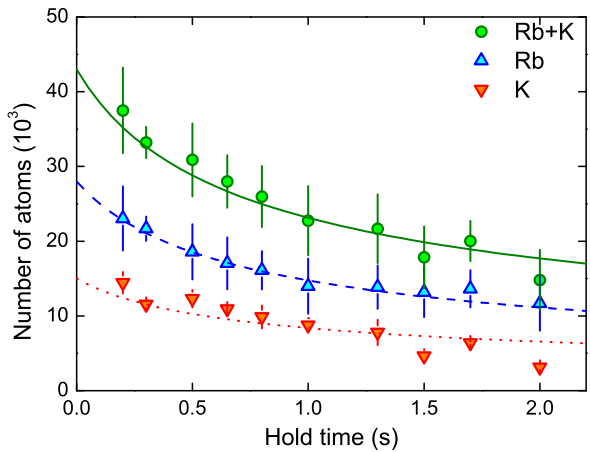}
  \caption{(Color online) Decay of the number of atoms trapped in our
    crossed dipole trap at the resonant magnetic field $B =
    56.8\,\text{G}$. At this magnetic field only collisions in KRbRb
    channel are relevant: the collision rate $\alpha_{\text{KRbRb}}$
    equals $1.16(1)\times 10^{-22}\,$cm$^6$/s. The lines show the
    results of our numerical model including only KRbRb 3BR
    collisions.}
  \label{fig:figure2}
\end{figure}
We apply a uniform magnetic field (Feshbach field) and hold the
optical trap at constant depth while the atom number decays due to 3BR
losses. Then, we switch off the trap and separately image the falling
K and Rb clouds. An example of atom decay during the hold time is
shown in Fig.~\ref{fig:figure2}. We record the atom losses as we scan
the interspecies scattering length varying the Feshbach field. For the
best signal-to-noise ratio, we use the total atom number $N(t_{\rm h})
= \Nk(t_{\rm h}) + \Nrb(t_{\rm h})$ after a fixed hold time $t_{\rm
  h}$ as our main observable, while the atom numbers of individual
species, $\Nk$ and $\Nrb$ are used to ascertain the dominant channel
of three-body losses at the resonance peaks. We point out that, since
our losses are due to 3BR collisions, our observable $N(t_{\rm h})$
yields the same information as the 3BR rate, as far as the position
and the width of the Efimov resonances are concerned.

\begin{figure*}
  \includegraphics[width=.95\textwidth]{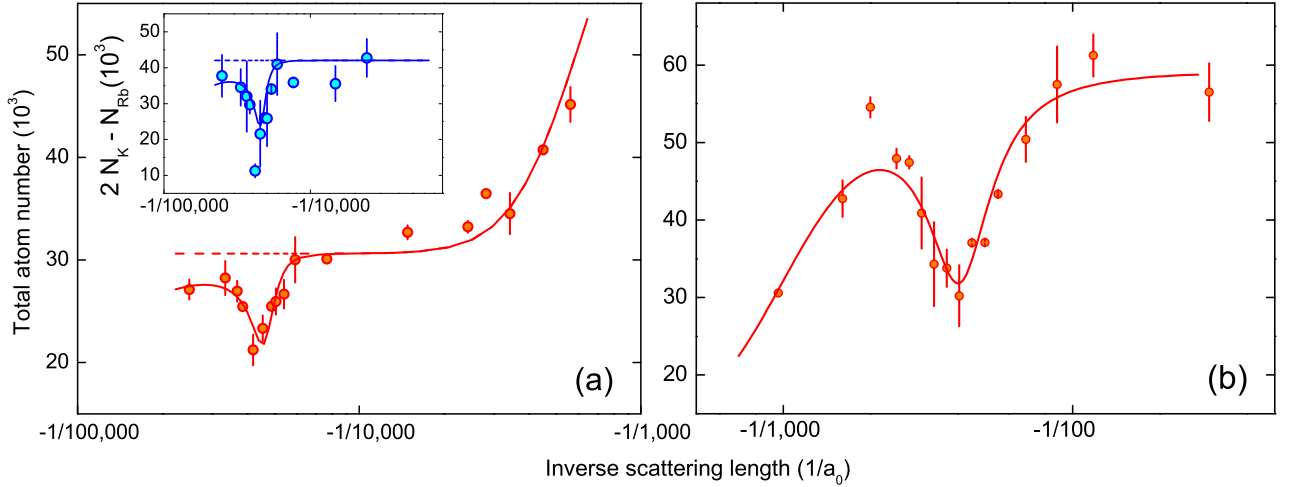}
  \caption{(Color online) Peaks of inelastic atomic losses at 38.8\,G
    (a) and 57.5\,G (b), signaling the Efimov resonances at negative
    scattering lengths in the KKRb and KRbRb channel, respectively. We
    show the total number of atoms $\Nk + \Nrb$ remaining in the
    optical trap after a fixed hold time at the given magnetic
    field. The hold times and temperatures are: 100\,ms and
    0.3\,$\mu$K (a), 500 ms and 0.4\,$\mu$K (b). The solid lines refer
    to the numerical results of our model. The dashed lines (a) show
    the result of numerical integration taking into account only KRbRb
    collisions; in the inset, we plot the linear combination $2\Nk -
    \Nrb$, which supports the assignment of the Efimov peak to KKRb
    collisions (see text).}
  \label{fig:figure3}
\end{figure*}
On the side of negative scattering length, i.e., for magnetic fields
above 38.4\,G, we observe two peaks of three-body losses above a
smooth increase with the scattering length (see
Fig.~\ref{fig:figure3}). The broadest peak, strong and visible up to
temperatures of approximately 0.8\,$\mu$K, is centered at a magnetic
field of 57.7(5)\,G. The second peak, weaker and visible only at our
lowest possible temperature of 0.3\,$\mu$K, lies at 38.8(1)\,G.
Conversion of these magnetic field values into scattering length is
done with the aid of the numerical results of the collisional model of
the KRb potential \cite{Simoni2008}, based on extensive studies of
Feshbach spectroscopy \cite{Ferlaino2006, Klempt2007} combined with
our recent spectroscopy measurements of the weakly bound molecular
state \cite{Weber2008}.

To make sure that the peaks are genuine interspecies 3-body features,
we have taken the following steps.  First, we verified that both loss
peaks are absent if we prepare samples with only a single
species. Second, according to the collisional model, at the
corresponding magnetic field values no sufficiently broad Feshbach
resonances occur, even considering molecular levels with angular
momentum $\ell=1,2$. Indeed, we do observe predicted narrow Feshbach
$d$-wave resonances, that further confirms the validity of the
collisional model \cite{NJP09-Thalhammer}. We also checked that the
frequency offset between the crossed trap beams is sufficiently far
detuned with respect to any bound state, in order to avoid Raman
transition to molecular levels.

Finally, at 56.8\,G we verified that the ratio of lost Rb to K atoms
is $1.7(3)$, which unambiguously proves that to the stronger resonance
peak is due to 3BR and it is dominated by the KRbRb channel. This is
shown by the time evolution of individual species atom number, see
Fig.~\ref{fig:figure3} .
For the weaker peak at 38.8 G, for best signal-to-noise we analyze the
individual species atom number at fixed hold time. From our dataset we
calculate the linear combination $2\Nk - \Nrb$, that is expected to be
constant for KRbRb collisions and to display a negative peak for KKRb
collisions. In the inset of Fig.~\ref{fig:figure3} we show that such a
peak is indeed observed and that experimental data for $2\Nk - \Nrb$
nicely agree with our numerical results. As a consequence we assign
the loss peak to the KKRb channel.

To derive the 3BR rates, we model the loss process by a set of
differential rate equations. We start from the local rate equations
for the atom densities $\nk$ and $\nrb$, i.e., $\dot \nk = -2\alphak
\nk^2\nrb -\alpharb\nk\nrb^2$ and $\dot \nrb = -\alphak\nk^2\nrb -2\alpharb\nk
\nrb^2 $ where $\alphak$ and $\alpharb$ denote the
event collision rates in the KKRb and KRbRb 3BR channels.
By integration upon coordinates, we obtain the rate equations for the
atom numbers $\Nk,\,\Nrb$. In addition, we also consider recombination
heating and evaporation, taking into account the displacement between
the two species due to the differential gravity sag. The set of (four)
differential equations, for atom number and total energy of each
species, is numerically integrated.
In analogy with the case of identical
particles, the 3BR rate $\alpharb$ is taken to be \cite{DIncao2006}:
$$
\alpharb=C\frac{\sinh
  2\eta}{\sin^2[s_0(\delta)\log(a/a_-)]+\sinh^2\eta}\frac{\hbar
  a_\delta^4}{\mu_\delta}
$$
with $\delta = \mk/\mrb$, $a_\delta =a\sqrt[4]{\delta(\delta
  +2)}/\sqrt{\delta + 1}$ and $\mu_\delta =
\mk/\sqrt{\delta(\delta+2)}$.
Likewise, we assume the equivalent expression for $\alphak$, with
independent $C$, $a_{-}$, $\eta$ parameters. The mass-dependent
scaling parameter $s_0(\delta)$ equals 0.644 for KRbRb and $0.246$ for
KKRb. The positions, widths and multiplicative factors ($a_-$, $\eta$,
$C$) are adjusted to match the numerical results of the above rate
equations with the measured number of atoms. For the weaker peak at
38.8\,G, it is crucial to introduce the unitary limit.  Indeed {\it
  for each channel} the 3BR rate is constrained by unitarity below the
theoretical upper limit \cite{DIncao2004} $\alpha_{\rm max} = 192
\pi^{2}\hbar/(\mu_\delta k^{4})$, that depends on temperature through
$(\hbar k)^{2}= 2\mu_{\delta}\kB T$. However, following the numerical
results of \cite{DIncao2004} we use as upper limit a value which is a
factor of 20 lower than the above expression. In practice, for each
channel we replace the appropriate $\alpha$, with an effective
$\alpha_{\rm eff} = 0.05 \alpha_{\rm max} \alpha/(0.05
\alpha_{\text{max}} + \alpha)$ approximately equal to the minimum
between $\alpha$ and $0.05\alpha_{\rm max}$.

The output of numerical integration, shown in Fig.~\ref{fig:figure3}
agree nicely with data, once we adjust the parameters to the
following values: $a_{-} = -246(14)\,a_0$, $\eta = 0.12(1)$ and 
$C =28(5)\times 10^3$ for $\alpharb$ and $a_- =
-22(^{+4}_{-6}) \times 10^3\,a_0$, $\eta = 0.02(1)$ and $C =
23(5)\times 10^{-4}$ for $\alphak$. 
The uncertainties on the positions reflect the estimated uncertainty
on the magnetic field values.

We find that the KRbRb channel dominate 3BR collisions at nearly all
magnetic fields, with the exception of a narrow region next to
Feshbach resonance where we detect the Efimov peak of the KKRb channel
(see dashed lines in Fig.~\ref{fig:figure3}a). It is important to note
that this peak is observable because, at these magnetic fields, the
dominant KRbRb channel is limited by unitarity, while the much weaker
KKRb channel is not.

\begin{figure}
  \includegraphics[width=.8\columnwidth]{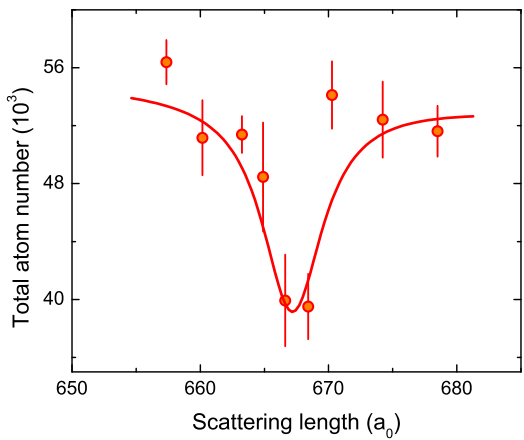}
  \caption{(Color online) Peak of enhanced inelastic atom losses at positive
    scattering lengths, corresponding to the atom-dimer resonance at 4.2\,G. We
    show the total number of atoms $\Nk +\Nrb$
    remaining in the optical trap after a fixed hold time of 500\,ms
    at the given the scattering length. These data were taken at a
    temperature of 400\,nK.  Each point averages on 2 to 5
    experimental runs.
  }
  \label{fig:figure4}
\end{figure}
For positive scattering length, the three-body inelastic collision
rate is expected to display an oscillatory behavior with no sharp
peaks. However, we observe a peak of atomic losses at $B =
4.25(0.10)$\,G, corresponding to $a = 667(1)\,a_0$. Loss peaks at
positive values of the scattering length have been recently reported
in \kt\ and attributed to resonantly enhanced secondary collisions
between atoms and dimers created in 3BR processes
\cite{Zaccanti2009}. 
The atom-dimer collisions are enhanced for values of the scattering
length, $a = a_\ast$, where the Efimov state intercepts the atom-dimer
threshold: at this values, a resonance occurs in the atom-dimer
scattering \cite{Knoop2008}, similar to a Feshbach resonance. In
analogy with \cite{Zaccanti2009}, the narrow peak of losses at 4.2\,G,
shown in Fig.~\ref{fig:figure4}, could be due to an atom-dimer
resonance. With the same arguments as above, we can exclude inelastic
two body collisions to be the cause of this peak.

Assuming the presence of an atom-dimer resonance, we modified the rate
equations to include the atom-dimer collisions using the following
formulas for the elastic cross-section %
$\sigma_{\rm AD}=C^\prime\times42.5 a_\delta^2/D$ and inelastic collision
rate $\beta_{\rm AD}=C^\prime\times 20.3 (\hbar
a_\delta/\mu_\delta)\sin(2\eta_\ast)/D$ where
$D=\sin^2[s_0\log(a/a_\ast)] +\sinh^2 \eta_\ast$ and $(C^\prime, \eta_{\ast}
a_{\ast})$ are free parameters. Since for this peak we lose more Rb
than K atoms, the resonance is assumed for the KRbRb channel only.

The fit results are: $C^\prime = 1.2(0.3)\times 10^{-5}$,
$\eta_{\ast} = 2(1)\times 10^{-3}$ and $a_{\ast} = 667(1)\,a_{0}$. We
notice that both the elastic cross section and the inelastic collision
rate are approximately five orders of magnitude lower than the
theoretical results for identical particles at zero temperature. Lower
than predicted values of the inelastic atom-dimer collision rate
$\beta_{\rm AD}$ have also been observed in Cs \cite{Knoop2008}. 
We remark, however, that a meaningful comparison of our data with
theoretical predictions will require, on one hand, the extension of
the homonuclear results to the heteronuclear case \cite{Helfrich2009},
on the other hand, the extension to finite temperature.  
If confirmed, the atom-dimer resonance would allow to assess the
energy of the Efimov state, that, as shown in the picture of
Fig.~\ref{fig:fig1}, is approximately equal to the dimer energy at
4.2\,G, i.e., $\sim h\times 0.2\,\text{MHz}$ \cite{Weber2008}.

In summary, we have observed three distinct peaks of the inelastic
collision rate of the mixture \kq\rb\ near an interspecies Feshbach
resonance. These peaks represent Efimov resonances, of both the KKRb
and KRbRb channel, occurring at values of the interspecies scattering
length $a$ such that the binding energy of Efimov trimers
vanishes. Our data represent the first unambiguous observation of
Efimov physics in systems composed of distinguishable particles with
different masses and the first experimental demonstration that two
resonant interactions are sufficient for Efimov physics to take
place. These findings have a direct impact on the search of Efimov
physics in a broad domain of physical systems, \emph{in primis}
nuclear physics.

We foresee that in the future more Efimov physics will emerge
especially with very asymmetric systems composed of one particle which
is much lighter than the other two: for such systems, the scaling
factor approaches 1 and several consecutive Efimov states could be
detected. With ultracold atoms, very promising combinations are \mbox{LiYbYb}
($e^{\pi/s_0}\simeq 4.5$), \mbox{LiCsCs} ($e^{\pi/s_0} \simeq 5.5$), as well
as \mbox{LiRbRb} ($e^{\pi/s_0} \simeq 7.9$) where an interspecies Feshbach
resonance has already been observed \cite{Deh2008}. Recently an
unexpected loss feature, perhaps an Efimov resonance, has been
observed in a system composed of fermionic atoms in three spin states
where all the scattering lengths are large \cite{Ottenstein2008}.

\begin{acknowledgments}
  We acknowledge fruitful discussions with M.~Zaccanti, G.~Modugno,
  J.~P.~D'Incao, G.~Poggi and P.~G.~Bizzeti. Funding was provided by
  CNR (EuroQUAM-DQS, EuroQUAM-QUDIPMOL), EU (STREP CHIMONO, NAMEQUAM)
  and INFN (SQUAT-Super).
\end{acknowledgments}

\bibliography{efimovprl-v2}
\end{document}